\documentclass[journal=jacsat,manuscript=article]{achemso}

\usepackage{chemformula} 
\usepackage[T1]{fontenc} 

\usepackage{graphicx} 
\usepackage[version=4]{mhchem} 
\usepackage[]{siunitx}

\author{Marijn Rikers}
\affiliation[Friedrich Schiller University Jena]{Institute of Solid State Physics, Friedrich Schiller University Jena, Max-Wien Platz 1, 07743, Jena, Germany}
\alsoaffiliation[Abbe Center of Photonics]{Institute of Applied Physics, Abbe Center of Photonics, Friedrich Schiller University Jena, Albert-Einstein-Straße 6, 07745, Jena, Germany}
\alsoaffiliation[Australian National University]{ARC Centre of Excellence for Transformative Meta-Optical Systems, Department of Quantum Science and Technology, Research School of Physics, Australian National University, 60 Mills Rd., Canberra, ACT 2601, Australia}
\email{marijn.rikers@uni-jena.de}

\author{Ayesheh Bashiri}
\affiliation[Friedrich Schiller University Jena]{Institute of Solid State Physics, Friedrich Schiller University Jena, Max-Wien Platz 1, 07743, Jena, Germany}
\alsoaffiliation[Abbe Center of Photonics]{Institute of Applied Physics, Abbe Center of Photonics, Friedrich Schiller University Jena, Albert-Einstein-Straße 6, 07745, Jena, Germany}

\author{Katsuya Tanaka}
\affiliation[Friedrich Schiller University Jena]{Institute of Solid State Physics, Friedrich Schiller University Jena, Max-Wien Platz 1, 07743, Jena, Germany}
\alsoaffiliation[Abbe Center of Photonics]{Institute of Applied Physics, Abbe Center of Photonics, Friedrich Schiller University Jena, Albert-Einstein-Straße 6, 07745, Jena, Germany}

\author{Ángela Barreda}
\affiliation[Friedrich Schiller University Jena]{Institute of Solid State Physics, Friedrich Schiller University Jena, Max-Wien Platz 1, 07743, Jena, Germany}
\alsoaffiliation[University Carlos III]{Department of Electronic Engineering, University Carlos III of Madrid, Avda. de la Universidad 30, 28911 Leganés, Spain}

\author{Duk-Yong Choi}
\affiliation[Australian National University]{ARC Centre of Excellence for Transformative Meta-Optical Systems, Department of Quantum Science and Technology, Research School of Physics, Australian National University, 60 Mills Rd., Canberra, ACT 2601, Australia}

\author{Thomas Pertsch}
\affiliation[Abbe Center of Photonics]{Institute of Applied Physics, Abbe Center of Photonics, Friedrich Schiller University Jena, Albert-Einstein-Straße 6, 07745, Jena, Germany}
\alsoaffiliation[Fraunhofer IOF]{Fraunhofer-Institute for Applied Optics and Precision Engineering IOF, Albert-Einstein-Str. 7, 07745 Jena, Germany}
\alsoaffiliation[Max Planck School of Photonics]{Max Planck School of Photonics, Hans-Knöll-Str. 1, 07745 Jena, Germany}

\author{Isabelle Staude}
\affiliation[Friedrich Schiller University Jena]{Institute of Solid State Physics, Friedrich Schiller University Jena, Max-Wien Platz 1, 07743, Jena, Germany}
\alsoaffiliation[Abbe Center of Photonics]{Institute of Applied Physics, Abbe Center of Photonics, Friedrich Schiller University Jena, Albert-Einstein-Straße 6, 07745, Jena, Germany}
\alsoaffiliation[Max Planck School of Photonics]{Max Planck School of Photonics, Hans-Knöll-Str. 1, 07745 Jena, Germany}

\title{Towards Polarization Routing of Magnetic and Electric Dipolar Emission with Dielectric Metasurfaces}

\abbreviations{}

\begin{document}

%

\begin{abstract}
We investigate the polarization properties of emission associated with the magnetic dipole and electric dipole transitions of europium(III) coupled to an anisotropic dielectric metasurface with polarization-engineered electric and magnetic photonic local density of states.
The metasurface consists of a square array of Mie-resonant elliptical a-Si:H dimers situated on an \ce{SiO$_2$} substrate and embedded in a PMMA film containing Eu(TTA)$_3$.
Based on reciprocity principle, it was designed to achieve maximum electric (magnetic) field enhancement in the dimer gap at 610\,nm (590\,nm) for $x$-polarized ($y$-polarized) normally incident light in order to selectively enhance the electric dipole (magnetic dipole) emission into the $x$-polarized ($y$-polarized) emission channel, respectively.
Momentum-resolved spectroscopy and back-focal plane imaging of emission of the fabricated light-emitting metasurface clearly reveal the intended polarization-dependent emission behaviour, with the $x$-polarized ($y$-polarized) emission showing a reduced (enhanced) ratio of the magnetic-/electric dipole emission intensity, correspondingly where the magnetic dipole emission is enhanced with a magnetic field enhancement from the nanostructures.
The demonstrated polarization-dependent interaction of a designed nanostructure with the electric- and magnetic dipolar transitions of trivalent lanthanide ions opens an avenue towards routing of emission of different multipolar orders into different polarization channels.
\end{abstract}

\section{Introduction}
Resonant dielectric metasurfaces have been established as a viable route to enhance and tailor spontaneous emission of integrated quantum emitters via the Purcell effect \cite{purcellSpontaneousEmissionProbabilities1946, liuLightEmittingMetasurfacesSimultaneous2018, yangOrchestratingSpontaneousEmission2025}.
Owing to their many spatial degrees of freedom they allow for tailoring the properties of spontaneously emitted light, including directionality, polarization, and, to some extent, wavefront \cite{vaskinLightemittingMetasurfaces2019,mohtashamiLightemittingMetalensesMetaaxicons2021,iyerUnidirectionalLuminescenceInGaN2020}. Since for most quantum emitters the strength of the \textit{electric dipole} (ED) transitions is several orders of magnitude larger than that of the \textit{magnetic dipole} (MD) transitions \cite{landauElectrodynamicsContinuousMedia2013}, research on light-emitting metasurfaces has so far mainly concentrated on the manipulation of spontaneous emission via ED transitions.
However, certain quantum emitters, such as trivalent lanthanide ions and semiconductor quantum dots, exhibit MD transitions with strengths equalling or even exceeding those of their ED transitions \cite{dodsonMagneticDipoleElectric2012,zurita-sanchezMultipolarInterbandAbsorption2002, carnallSpectralIntensitiesTrivalent1968,juddOpticalAbsorptionIntensities1962,ofeltIntensitiesCrystalSpectra1962}. 
MD transitions have received significant research attention due to their important application prospects, e.g. for atomic clocks, quantum memory and information processing, as well as sensing \cite{yudinMagneticDipoleTransitionsHighly2014, ahlefeldtQuantumProcessingEnsembles2020, khalilEuropiumBetadiketonateTemperature2004}.
In analogy to ED emitters, the spontaneous emission rate of an MD emitter changes when it is placed into a specific photonic environment, as described by the Purcell effect.
Formally, the electric and magnetic Purcell effects follow equivalent expressions for the modified emission rate, which were already expanded and discussed elsewhere (see e.g. \cite{baranovModifyingMagneticDipole2017}).
In these expressions, the influence of the environment enters through the electric (magnetic) local density of states (LDOS), respectively.
Importantly, the electric and magnetic Purcell factors, $F_P^{\text{el, mag}}$, are defined as the ratio between the spontaneous emission rate $\gamma$ of an emitter in a given environment and the corresponding emission rate $\gamma_0$
in free space. These quantities depend exclusively on the electromagnetic properties of the surrounding medium and are determined by the imaginary part of the appropriate Green’s tensor associated with electric or magnetic dipole sources \cite{baranovModifyingMagneticDipole2017}.
While Purcell factors can, in principle be calculated numerically for almost arbitrary photonic environments, these calculations are computationally expensive for metasurfaces, which are typically characterized by periodic boundary conditions and anisotropic environments.
An alternative effective way to quantitatively describe the change of spontaneous emission rate for both ED and MD emitters integrated into metasurfaces is based on the reciprocity principle \cite{vaskinLightemittingMetasurfaces2019, vaskinManipulationMagneticDipole2019, bashiriColorRoutingEmission2024}.
Thereby, numerical models, that explicitly consider localized dipole sources, are replaced by the calculation of electromagnetic near-fields within the metasurface unit cell  under far-field plane wave excitation. Under these conditions, large electric (magnetic) near-field enhancements at the emitter position are directly associated with high electric (magnetic) Purcell factors.

A variety of nanophotonic structures has so far been demonstrated to allow for the manipulation of MD emission via the Purcell effect \cite{ hussainEnhancingEu3Magnetic2015,sanz-pazEnhancingMagneticLight2018,vaskinManipulationMagneticDipole2019,sugimotoMagneticPurcellEnhancement2021,bashiriColorRoutingEmission2024}.
Most of these works rely on trivalent europium (\ce{Eu^3+}) as emitting species. With its bright red color, \ce{Eu^3+} is widely used in many applications, with fluorescent lighting being the most eminent example \cite{smetsPhosphorsBasedRareearths1987, boddulaWhiteLightemissiveEuropium2021}.
Its emission arises from several bright optical transitions in the visible range, including a MD transition at $\lambda_\mathrm{em} \approx \SI{590}{\nano\meter}$ (\ce{{}^5D0\bond{->}{}^7F1}) and an ED transition at $\lambda_\mathrm{em} \approx \SI{610}{\nano\meter}$ (\ce{{}^5D0\bond{->}{}^7F2}), among other (\ce{{}^5D0\bond{->}{}^7F_i}) transitions $(i\in\{0-6\})$ \cite{binnemansInterpretationEuropiumIIISpectra2015, thorEuropiumIIICoordinationChemistry2024}.
The existence of MD transitions in \ce{Eu^3+} was first experimentally demonstrated by Drexhage, who studied europium-containing complexes positioned at variable distances from a dielectric interface \cite{drexhageInfluenceDielectricInterface1970}. Other experimental configurations were also used to demonstrate the MD response of \ce{Eu^3+}, most prominently momentum-resolved spectroscopy \cite{taminiauQuantifyingMagneticNature2012}. 

Several works have studied \ce{Eu^{3+}} emission in tailored nanophotonic environments, including plasmonic metasurfaces, dielectric nanoantennas, dielectric metasurfaces, and dielectric colloidal particles \cite{hussainEnhancingEu3Magnetic2015,sanz-pazEnhancingMagneticLight2018,vaskinManipulationMagneticDipole2019,sugimotoMagneticPurcellEnhancement2021}.
A typical observable is the modification of the detected ratio $G = I_{\mathrm{MD}}/I_{\mathrm{ED}}$ or similar, where $I_{\mathrm{ED}}$ $(I_{\mathrm{MD}})$ is the emission intensity of the ED (MD) transition.
Directional control of the MD transition was also demonstrated by Bashiri \textit{et al.}, with the MD emission being mainly routed into the normal direction, while the ED emission was routed over larger angles by utilizing a metasurface exhibiting quasi-bound states in the continuum \cite{bashiriColorRoutingEmission2024}.
The demonstrated selective interaction with ED and MD transitions is relevant for various applications.
For instance, selectively enhancing and controlling specific transitions enables more efficient qubit operations and extended coherence times in quantum information processing \cite{zhongOpticallyAddressableNuclear2015}, while isolating and enhancing targeted transitions improves sensitivity and selectivity in sensing applications \cite{parkerExcitementBlockStructure2004}.
 However, to achieve the desired selectivity needed to realize the routing effect, Bashiri \textit{et al.} relied on the difference in wavelength between the ED and MD transitions \ce{Eu^{3+}} in combination with spectrally narrow, high-quality (Q) factor resonances.
Consequently, the approach is limited to spectrally well separated ED and MD transitions. Moreover, structural imperfections and material losses can easily degrade the performance of high-Q structures, rendering their fabrication highly demanding. Also, while directional routing is of interest for microscopy-based applications, many other application areas, including e.g. telecommunications \cite{fatomeObservationLightbylightPolarization2010} or quantum optics \cite{tangExperimentalDemonstrationPolarization2014}, favour polarization as spatial degree of freedom for information encoding.
However, routing of light emitted via the ED and MD transitions of \ce{Eu^{3+}} into orthogonal far-field polarizations has not been attempted so far. 
By reciprocity, a photonic nanostructure that provides both electric and magnetic Purcell enhancement and is simultaneously able to route the ED and MD emission into different far-field polarizations, needs to be anisotropic and exhibit electric and magnetic near-field enhancement for excitation with plane waves of orthogonal linear polarizations, respectively.
A prototypical structure for the selective enhancement of electric and magnetic near-fields under plane-wave excitation are dielectric dimers composed of two Mie-resonant nanoresonators \cite{mieBeitrageZurOptik1908,evlyukhinDemonstrationMagneticDipole2012} separated by a small feed-gap\cite{bakkerMagneticElectricHotspots2015}.
In addition to the common electric hot-spots, which are well studied for plasmonic feedgap geometries \cite{kinkhabwalaLargeSinglemoleculeFluorescence2009}, such dielectric dimers can also feature hot-spots of the magnetic field in the feed-gap.
Importantly, the electric and magnetic hot-spots are supported by two different modes, which can be selectively excited by incident light of orthogonal linear polarizations \cite{ bakkerMagneticElectricHotspots2015}.
As such, the electric hot-spot occurs for normally incident light with the electric-field vector oriented along the dimer axis, while the magnetic hot-spot is excited for normally incident light with the electric-field vector oriented normal to that axis.
While there are many challenges connected to experiments that study single dimers, the described polarization-dependent response is in essence inherited by dimer arrays which are experimentally more viable \cite{zhangChiralEmissionResonant2022}. 

\begin{figure}[hbt!]
	\centering
	\includegraphics[width = 0.9\textwidth]{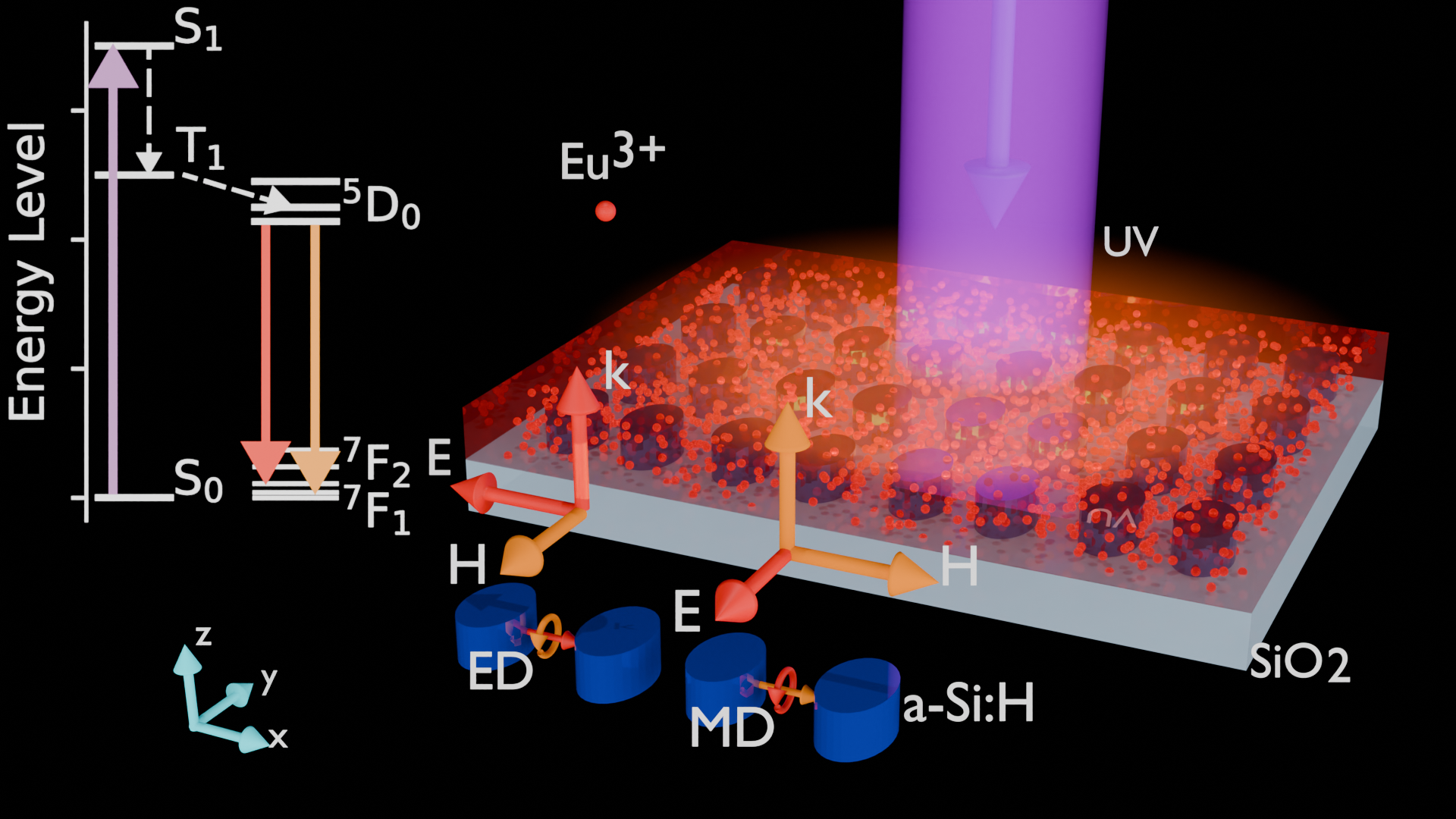}
	\caption{Concept of polarization routing metasurfaces. In the image, we see the a-Si:H dimer array metasurface encapsulated in a layer of \ce{PMMA$:$Eu(TTA)_3} where the red dots represent the \ce{Eu(TTA)_3} emitters are excited by unpolarized UV light and start to fluoresce. In the top left corner the energy diagram of \ce{Eu(TTA)_3} is shown, the upward purple arrow shows the UV excitation, the white dashed arrows indicate non radiate decay process, and the red and orange downward arrows indicate the ED and MD transitions, respectively. In the for ground the working principle of the dimers is shown, where EDs oriented along the dimer axis $(x)$ are enhanced by electric field enhancement, at the ED transition wavelength \SI{610}{\nano\meter}, resulting in $x$-polarized emission. While simultaneously MDs oriented along the dimer axis are enhanced by magnetic field enhancement, at the MD transition wavelength \SI{590}{\nano\meter}, resulting in $y$-polarized emission.}
	\label{fig:Concept}
\end{figure}

Here we leverage the polarization-dependent near-field properties of low Q-factor Mie-resonant dielectric dimers to enhance the emission of \ce{Eu^{3+}} into distinct polarizations depending on the dipolar nature, ED or MD, of the respective transition, as conceptually illustrated in Fig.~\ref{fig:Concept}.
Specifically, we arrange dielectric dimers composed of elliptical silicon nanoresonators in a square array and hybridize them with a with a PMMA film containing \ce{Eu^{3+}} to form anisotropic, light-emitting metasurfaces with engineered polarization properties.
Using momentum-resolved spectroscopy and back-focal plane (BFP) imaging of emission of the fabricated structures, we experimentally demonstrate the intended reduction (enhancement) of the ratio of the MD/ED emission intensity for the $x$-polarized ($y$-polarized) emission channel, correspondingly.
As such, making use of the selective interaction of designed nanostructures with electronic transitions of different multipolar order, our work paves the way towards polarization routing of different multipolar content of emitted radiation, with potential applications in communications, sensing, and quantum optics. 

\section{Design and Sample Preparation}
\begin{figure}[hbt!]
	\centering
    \includegraphics{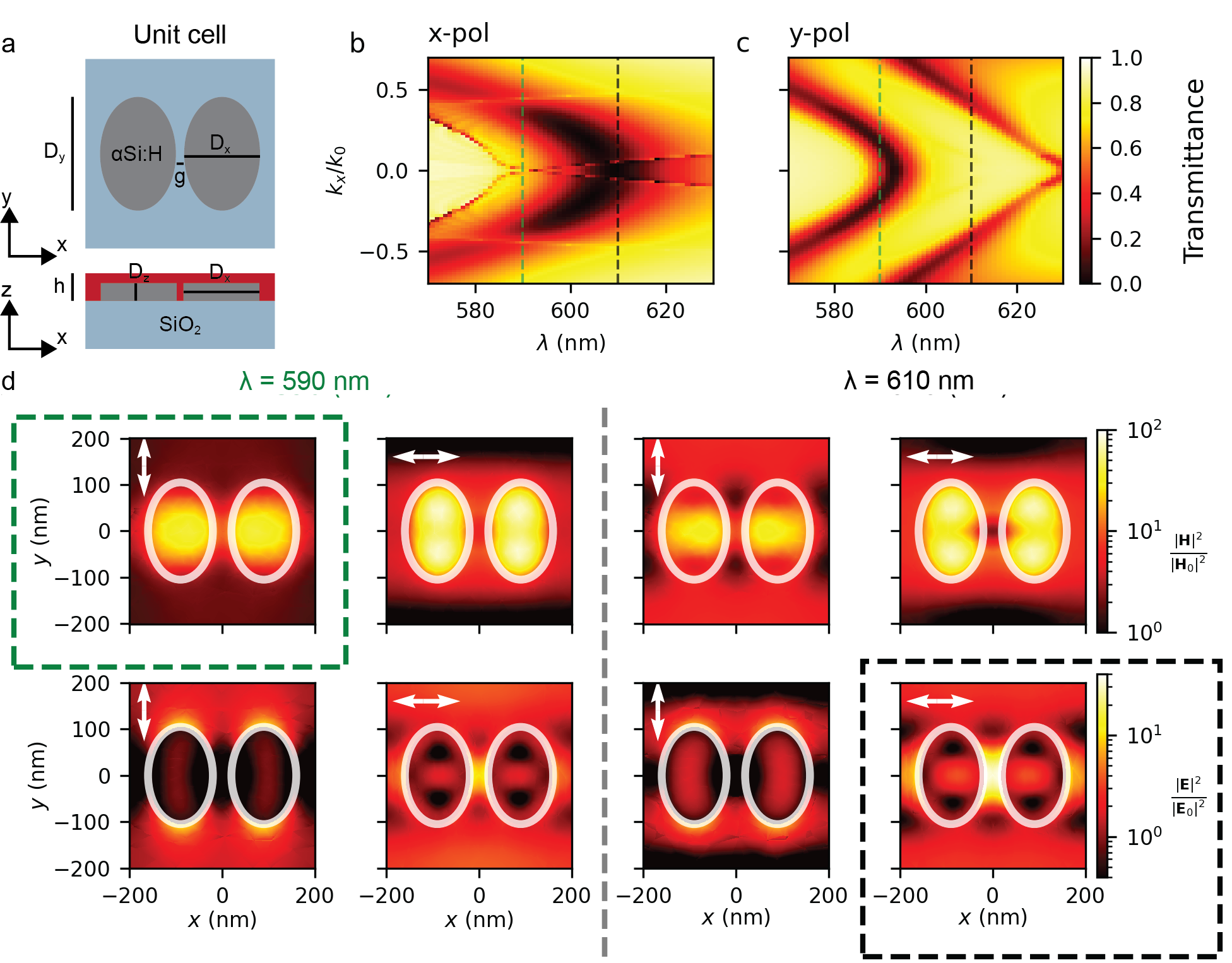}
    \caption{a) Schematic of the dimer array unit cell showing elliptical dimers with parameters $D_x = \SI{120}{\nano\meter}$, $D_y = \SI{180}{\nano\meter}$, $D_z = \SI{80}{\nano\meter}$, and $g = \SI{40}{\nano\meter}$, covered with a $h = \SI{150}{\nano\meter}$ thick PMMA layer. b,~c) Simulated linear momentum resolved transmission spectra of the metasurface for (b) $x$- (c) $y$-polarized incident light. The green dashed line indicates \SI{590}{\nano\meter} and the black dashed line indicates \SI{610}{\nano\meter}. d) Calculated near-field intensity profiles in the $x-y$ plane through the center of the nanoresonators. The top row shows the magnetic, the bottom row the electric field intensity enhancement. The polarization direction is indicated by the white arrow in each color map. The left half are maps at $\lambda = \SI{590}{\nano\meter}$ and the right half are maps at $\lambda = \SI{610}{\nano\meter}$ separated by the gray dashed line. The green and black dashed boxes highlight the cases of most pronounced electric and magnetic hotspot formation in the feedgap, respectively.}
    \label{fig:Simmulations}
\end{figure}

As a first step, we performed full-wave numerical calculations using the commercial software package Lumerical to design a dielectric dimer metasurface featuring hot-spots of the electric (magnetic) field at the ED (MD) transition wavelength of the \ce{Eu^3+}, respectively.
Each unit cell consists of two silicon nanoresonators arranged in close proximity to each other along the $x$-axis to form a dimer (see Fig.~\ref{fig:Simmulations}\,(a) for a sketch).
For the silicon dispersion we use experimental data obtained from ellipsometry measurements of an unstructured thin-film (see S2 for more details).
The dimers are situated on a glass substrate with a refractive index of $n_{\text{glass}}$ of 1.46 and embedded into a PMMA layer (refractive index $n_{\text{PMMA}}=1.49$) with a thickness of $h_{\text{PMMA}}=150$ nm.
Bloch periodic boundary conditions are employed in $x$- and $y$- direction, in $z$- direction perfectly matched layer boundary conditions are used.
Further details on the numerical model are summarized in the Methods Section.
To increase the number of degrees of freedom available for tailoring the optical modes as compared to classical disk based resonators, we chose elliptical nanoresonators as basic building blocks for the dimers \cite{bakkerMagneticElectricHotspots2015, rikersPolarizationDependentEnhancement2025}.
Each such nanoresonator provides three independent geometric degrees of freedom: the height $D_z$, the short-axis diameter $D_x$, and the long-axis diameter $D_y$, or, equivalently, the aspect ratio $\alpha$ defined by $D_y = \alpha D_x$.
In combination with the gap size $g$ and the period $p$, these parameters enable substantial control over the resonance wavelengths and their polarization characteristics.
Depending on the polarization of the incident electric field, $x$-polarized or $y$-polarized, an electric or magnetic hotspot is generated, respectively.
When emitters are located in these hotspots, their transition rates are increased because of the increase in the local density of optical states (LDOS) \cite{taminiauQuantifyingMagneticNature2012}. 

We conducted a parameter sweep to optimize the structure geometry for maximum electric (magnetic) field enhancement at a wavelength of \SI{610}{\nano\meter}  (\SI{590}{\nano\meter}) for $x$- ($y$-) polarized normally incident light, respectively. 
The period of the metasurface was fixed to $p=\SI{400}{\nano\meter}$ and the feedgap size to $g = \SI{40}{\nano\meter}$ to account for the finite resolution of the fabrication process. 
The optimized nanoresonator parameters were found as $D_x = \SI{120}{\nano\meter}$, $D_y = \SI{180}{\nano\meter}$, and $D_z = \SI{80}{\nano\meter}$. Fig.~\ref{fig:Simmulations}\,(b,~c) shows calculated transmission spectra of the optimized metasurface for $x$- and $y$-polarized incident light as a function of the normalized transverse momentum along the $x$-axis, $k_x/k_0$.
We observe two dispersive resonant features at the $\Gamma$ point ($k_x/k_0 = 0$) at \SI{610}{\nano\meter} (\SI{590}{\nano\meter}) wavelength for $x$- and $y$-polarized incident light, correspondingly. 
The near-field intensity profiles associated with these resonances are displayed in Fig.~\ref{fig:Simmulations}\,(d),  showing the formation of an electric and a magnetic field hotspot in the feedgap of the dimer for the relevant combinations of wavelength and incident polarization (highlighted by green and black frames in Fig.~\ref{fig:Simmulations}\,(d)). 
While the enhancement is the strongest in the feedgap, significant enhancement is also observed when averaging over the entire \ce{PMMA} volume. 
Also note that the $\mathbf{E}$ and $\mathbf{H}$ fields are also enhanced for the other two combinations of incident polarization and wavelength, however, they are mainly localized inside the resonators ($x$-polarized, \SI{590}{\nano\meter}) or at the tips of the resonators ($y$-polarization, \SI{610}{\nano\meter}).

To quantify performance, we computed the volume-averaged field enhancement factors (electric or magnetic) relative to a bare \ce{PMMA} film of identical thickness. The results, together with the polarization contrast values, are summarized in Table \ref{tab:Fp}. The polarization contrast measures how strongly the target field is enhanced under one polarization relative to the orthogonal one (positive values indicate selectivity for $y$-pol, negative for $x$-pol). The design achieves high contrast for the desired cases while suppressing unwanted cross-polarization response---in particular, minimizing electric enhancement under $y$-pol at \SI{610}{\nano\meter}. By reciprocity, these enhancements correspond to the electric (magnetic) Purcell factors, which govern the radiative decay rates of electric (magnetic) dipole emitters and thus enable polarization-selective spontaneous emission control \cite{baranovModifyingMagneticDipole2017}. The formal definition of polarization contrast and the composite figure of merit (FOM) used for optimization are provided in the Methods section.

Next, we fabricated the dimer metasurfaces targeting the optimized geometrical parameters.
To this end, hydrogenated amorphous silicon \ce{$\alpha$-Si$:$H} films with a thickness of $h = \SI{80}{\nano\meter}$ were nanostructured using a combination of electron-beam lithography and inductively coupled plasma etching.
To account for expected fabrication tolerances, we fabricated a series of metasurfaces with a variation of the geometrical design parameters in the layout file.
Specifically, the exposure included dimers with nominal dimensions of $(D_x = \SI{140}{\nano\meter}, D_y = \SI{210}{\nano\meter})$, $(D_x = \SI{150}{\nano\meter}, D_y = \SI{225}{\nano\meter})$, and $(D_x = \SI{160}{\nano\meter},  D_y = \SI{240}{\nano\meter})$, with $g = \SI{40}{\nano\meter}$ and $p=\SI{400}{\nano\meter}$, and a series $(D_x = \SI{140}{\nano\meter}, D_y = \SI{210}{\nano\meter})$, $(D_x = \SI{150}{\nano\meter}, D_y = \SI{225}{\nano\meter})$, and $(D_x = \SI{160}{\nano\meter},  D_y = \SI{240}{\nano\meter})$, with $g = \SI{60}{\nano\meter}$ and $p=\SI{440}{\nano\meter}$.
All metasurfaces were fabricated with a footprint of $100\times100$ unit cells. After fabrication, the sample was covered in a \SI{150}{\nano\meter} layer of \ce{PMMA$:$Eu(TTA)_3}, a \ce{PMMA} resist containing the europium complex \ce{Eu(TTA)_3} at a concentration of \SI{0.16}{\micro\mol\per\ml}. More details on the fabrication process are included in the S1.

\begin{table}[ht]
  \centering
  \begin{tabular}{@{} c||cccc @{}}
    \hline
    $\lambda\,(\unit{\nano\meter})$  & \multicolumn{2}{c}{$590$} & \multicolumn{2}{c}{$610$}  \\
    \hline\hline
    pol             & x     & y  & x & y \\ \cline{3-3}
    $\frac{|\mathbf{H}|^2}{|\mathbf{H}_0|^2}$         & 4.0   & \multicolumn{1}{|c|}{7.8} & 4.7 & 5.9  \\  \cline{3-4}
    $\frac{|\mathbf{E}|^2}{|\mathbf{E}_0|^2}$         &  2.3    & 1.6  &\multicolumn{1}{|c|}{3.2}& 1.1  \\ \cline{4-4}
  \end{tabular}
  \caption{Calculated integrated field enhancement. The row $\lambda$ indicates the wavelength, (pol) indicates the polarization, $\frac{|\mathbf{H}|^2}{|\mathbf{H}_0|^2}$ and $\frac{|\mathbf{E}|^2}{|\mathbf{E}_0|^2}$ are the electric and magnetic enhancement factors, respectively.}
  \label{tab:Fp}
\end{table}
 
\section{Results and Discussion}
\begin{figure}[hbt!]
	\centering
        \includegraphics[width = 0.9 \textwidth]{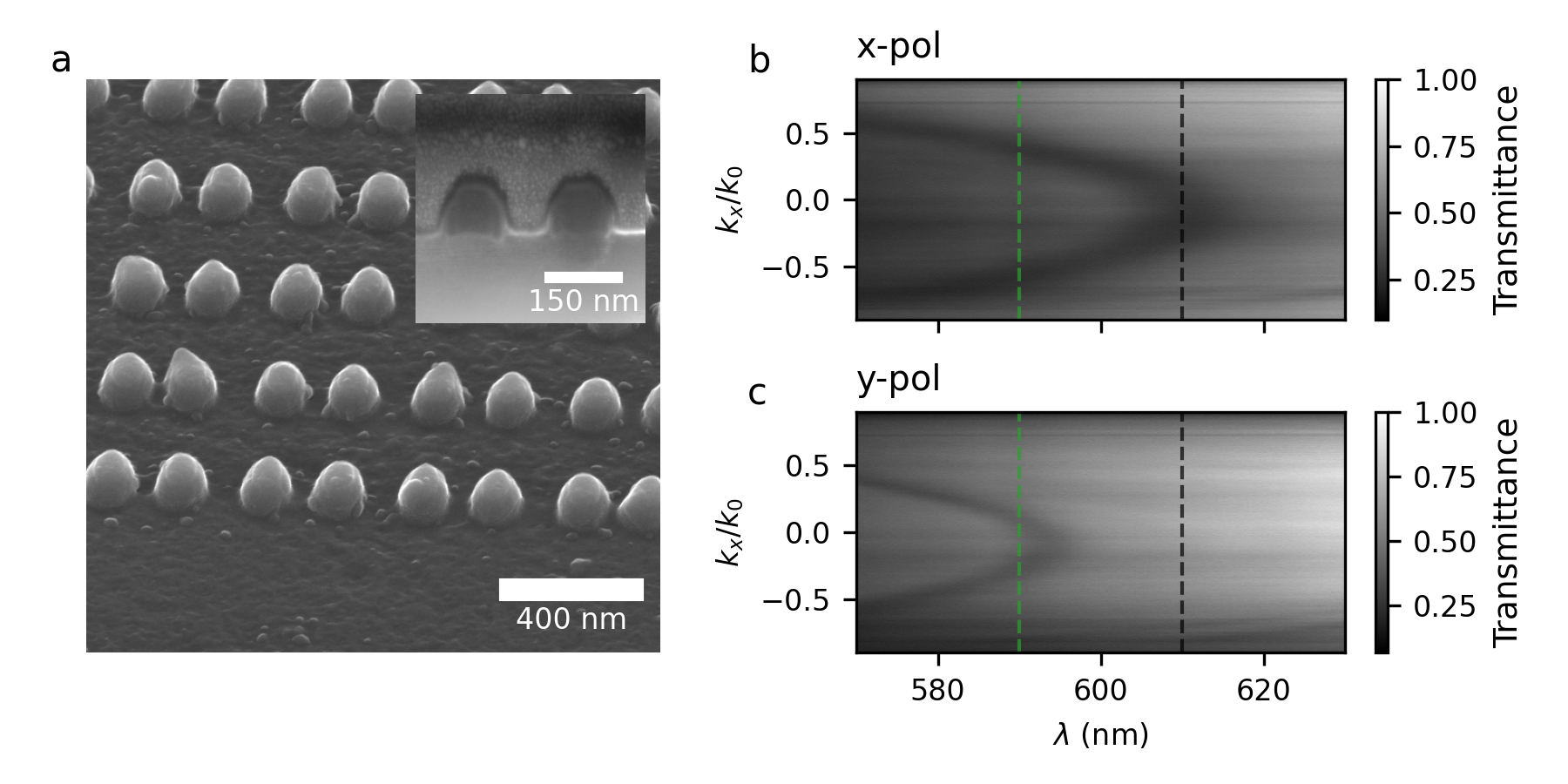}
    	\caption{a) Oblique-view (\SI{52}{\degree}) scanning-electron micrograph (SEM) of the fabricated metasurface. The inset shows an SEM of a focused-ion-beam (FIB) cross section of the sample. (b,~c) Momentum-resolved linear-optical transmission spectra of the metasurface shown in (b) for $x$- and (c) $y$-polarized incident light. Here the data is normalized to the  maximum transmittance. The green dashed line and black dashed line correspond to the MD (\SI{590}{\nano\meter}) and ED (\SI{610}{\nano\meter}) transition wavelength, respectively.}
    	\label{fig:SupportCaracter}
\end{figure}

 A scanning-electron micrograph (SEM) of the fabricated metasurface before application of the \ce{PMMA$:$Eu(TTA)_3} layer is shown in Fig.~\ref{fig:SupportCaracter}\,(a).
 The inset shows an SEM of a cross-section of the sample prepared using focused-ion-beam (FIB) milling.
 We observe that the side-walls of the nanoresonators are tilted, leading to deviations from the designed shape.
 Moreover, the FIB-cut reveals a thin oxidation layer on the a-Si:H resonators. The structure dimensions retrieved from the SEM inspection are $D_x = \SI{140}{\nano\meter}$, $D_y = \SI{200}{\nano\meter}$, $h=\SI{90}{\nano\meter}$, $p=\SI{440}{\nano\meter}$, and $g=\SI{80}{\nano\meter}$, with an uncertainty of \SI{5}{\nano\meter}.
 The lateral nanoresonator dimensions $D_x$ and $D_y$ are measured at their bottom and are larger than the design parameters by approximately \SI{20}{\nano\meter}.
 The nanoresonator height $D_z$ is increased by \SI{10}{\nano\meter}.
 Effectively, these increased dimensions compensate for the resonance blue-shift associated with the tilted sidewalls and oxidation layer (see discussion of measured optical properties).
 The gap size $g$ is increased by \SI{20}{\nano\meter}. 
Next, to examine the optical mode structure of the fabricated metasurface including the \ce{PMMA$:$Eu(TTA)_3} layer, we performed momentum-resolved linear-optical transmission spectroscopy.
To this end, we used a custom-built optical setup in transmission geometry with a tungsten halogen lamp as light source.
The BFP of the collection objective is projected onto the entrance slit of an imaging spectrometer using a 4f system (see S3 for details).
Fig.~\ref{fig:SupportCaracter}\,(b, c) shows the measured transmission spectra for (b) $x$- and (c) $y$-polarized light as a function of $k_x/k_0$.
In good agreement with numerical calculations, we clearly observe two dispersive resonant features with an apex at the $\Gamma$ point $(k_x/k_0 = 0)$ at \SI{610}{\nano\meter} and \SI{590}{\nano\meter} wavelength for $x$- and $y$-polarized incident light, respectively. 
The measured results show that the variations in nanoresonator shape and dimensions effectively compensate each other, all together resulting in an optical response exhibiting all the intended features.

\begin{figure}[hbt!]
	\centering
    	\includegraphics[width = 0.9\textwidth]{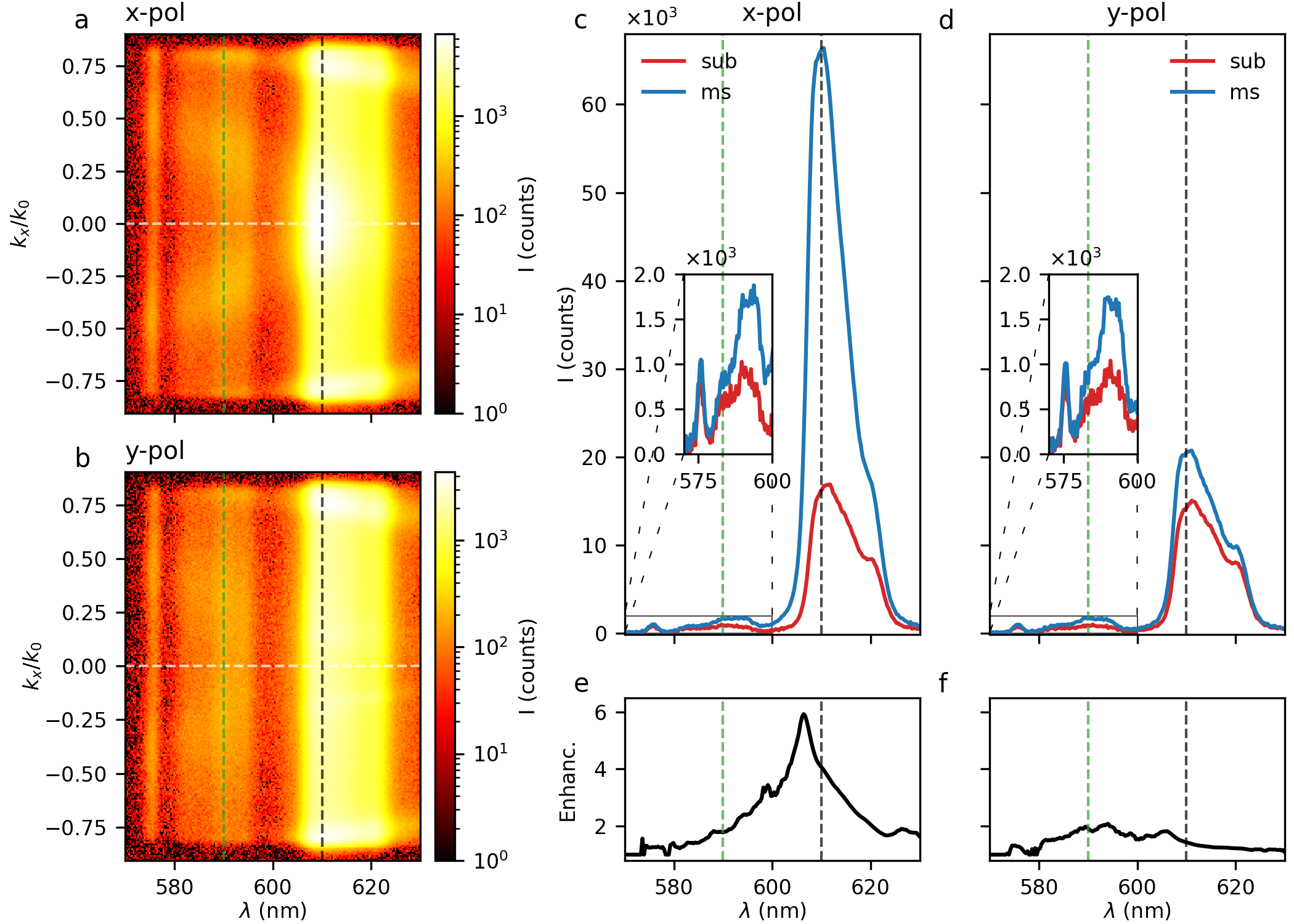}
    	 \caption{Momentum-resolved fluorescence spectra of the hybrid metasurface for (a) $x$- and (b) $y$- polarized detection. (c, d) Fluorescence spectra averaged around the $\Gamma$ point ($(k_{x}/k_0 = [-0.035,+0.035]$) for the hybrid metasurface and the same \ce{PMMA$:$Eu(TTA)$_3$} film on the bare substrate, labeled as ms and sub in the legend, respectively. Here (c, d)) shows the $x$-polarized ($y$-polarized) emission. The insets in (c, d) show a the data between $\lambda = \SI{570}{\nano\meter}$ and \SI{610}{\nano\meter} magnified. (e, f) Fluorescence enhancement spectra calculated by dividing the metasurface spectra by the substrate spectra for $x$- and $y$-polarized detection, respectively. The green and black vertical dashed lines incidence the MD and ED transition locations, respectively.}
    	\label{fig:PlResult}
\end{figure}

Next, we studied the polarization-dependent fluorescence properties of the hybrid metasurface using the same home-built setup as used for transmission measurements, however, in a different configuration.
To optically excite the \ce{Eu(TTA)3}, we illuminated the sample using an incoherent and unpolarized LED light source with a wavelength of \SI{365}{\nano\meter}.
The sample fluorescence was collected and analyzed in reflection geometry regarding its polarization, spectral and spatial properties. Further measurement details, including a sketch of the employed setup, are included in the Supporting Information.
Measured momentum-resolved fluorescence spectra for $x$- and $y$-polarized detection are shown in Fig.~\ref{fig:PlResult}\,(a,~b), respectively. 
The fluorescent lines of the \ce{Eu(TTA)_3} transitions \ce{{}^5D_0\bond{->}{}^7F_i} $i\in[0,1,2]$ occurring at $\lambda = [\SI{580}{\nano\meter},\SI{590}{\nano\meter},\SI{610}{\nano\meter}]$, appear as vertical lines.
Whenever one of the metasurface modes overlaps with an emission line, we observe an enhancement of the detected fluorescence signal, allowing us to trace their dispersion as bright lines in the momentum-resolved spectra.
In particular, at the $\Gamma$-point clear enhancement is observed for $x$-polarization at $\lambda = \SI{610}{\nano\meter}$, and for $y$-polarization at $\lambda = \SI{590}{\nano\meter}$.
For a quantitative analysis, Fig.~\ref{fig:PlResult}\,(c,~d)  
show the fluorescence of the hybrid metasurface and of the same \ce{PMMA$:$Eu(TTA)_3} film on the bare substrate, averaged over a small region of $k_x/k_0 = [-0.035, +0.035]$ around the Gamma point for $x$-polarization and $y$-polarization, respectively.
The $x$-polarized ED transition has the highest intensity.
The insets show a close up of the fluorescence spectra around the MD transitions and show a significant change in the spectral peak when comparing the $x$-polarized and $y$-polarized spectra.
Fig.~\ref{fig:PlResult}\,(e,~f) shows the fluorescence enhancement, calculated as the detected intensity from the metasurface divided by that of the same \ce{PMMA$:$Eu(TTA)_3} film on the bare substrate, averaged over a small region of $k_x/k_0 = [-0.035, +0.035]$ around the $\Gamma$ point for $x$- and $y$- polarization channels, respectively.
Most prominently, for the $x$-polarized emission we observe an enhancement of $5$ at $\lambda = \SI{610}{nm}$ with the maximum enhancement reaching $6$ at \SI{606}{\nano\meter}. At $I(\lambda = \SI{590}{\nano\meter})$, i.e. around the magnetic dipole transition of the \ce{Eu(TTA)_3}, the enhancement is about two fold, composed of a strong, spectrally broad contribution from the tail of the neighboring transition and a smaller contribution from the narrow emission peak of the MD transition.
For $y$-polarization, the fluorescence at $I(\lambda = \SI{590}{\nano\meter})$ is also enhanced by a factor of two, however, here the dominant contribution stems from the narrow emission peak of the MD transition.
The enhancement of the ED transitions is significantly smaller than for $x$-polarized detection, reaching only $1.4$ at $\lambda = \SI{610}{nm}$.
This represents a significant increase in brightness for a transition that is typically used as a calibration reference because of its robustness against environmental changes \cite{thorEuropiumIIICoordinationChemistry2024}. 
Importantly, the differences in enhancement of the dominant electric and the magnetic dipole transitions lead to a pronounced difference in the $G=I(\text{MD})/I(\text{ED})$ ratio for both polarizations, which we determine as $G^{\text{ms}}_{x-\text{pol}} = 0.03$ and $G^{\text{ms}}_{y-\text{pol}} = 0.09$, indicating a polarization-dependent change of $\sim3\times$ on the metasurface.
For reference, the value for the \ce{PMMA$:$EU(TTA)_3} film on the bare substrate is $G^{\text{sub}}_{x-\text{pol}} = G^{\text{sub}}_{y-\text{pol}} = 0.06$. 

\begin{figure}[hbt!]
	\centering
    	\includegraphics[width = 0.9\textwidth]{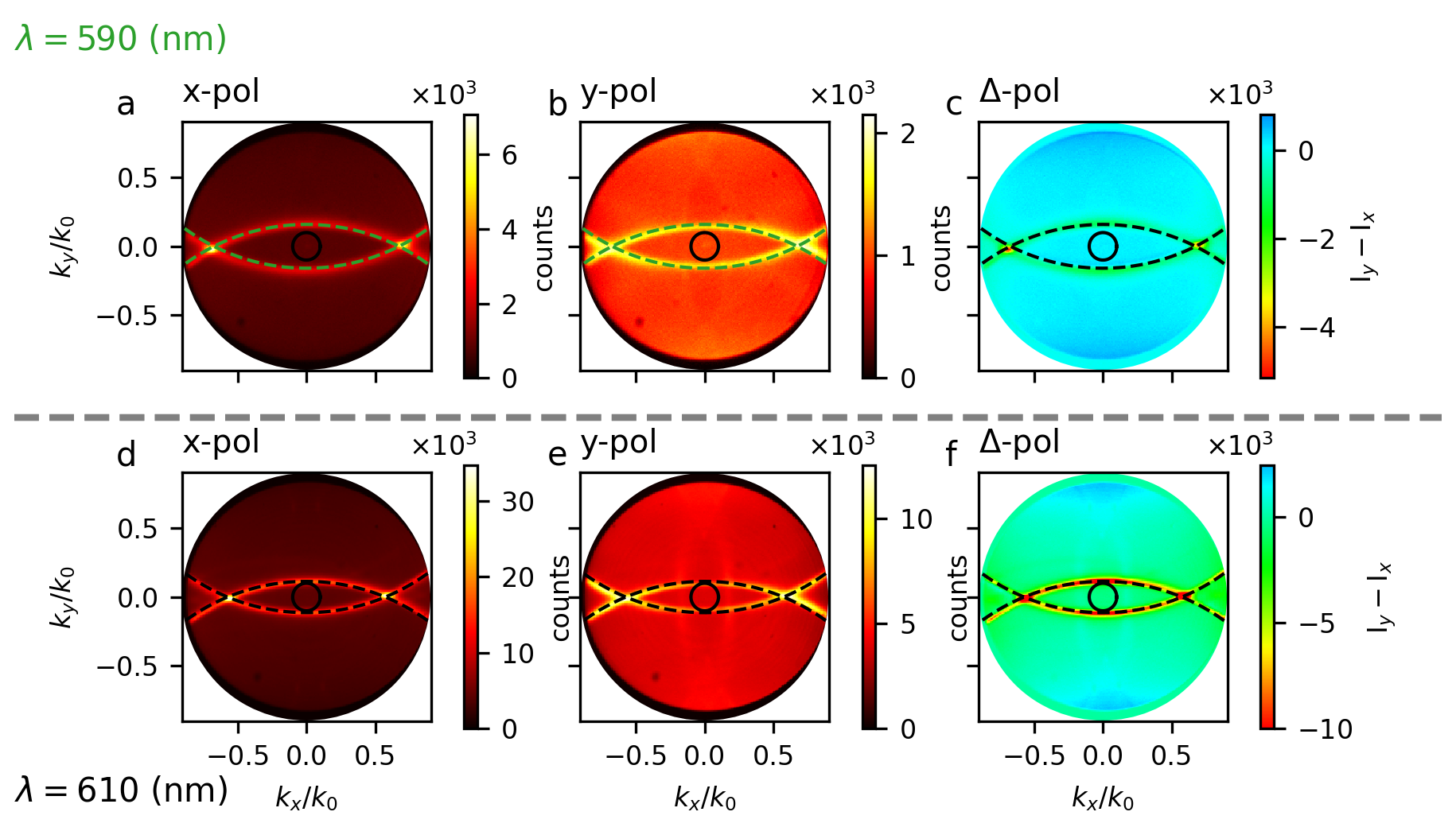}
	\caption{Back-focal plane images of the fluorescence from the dimer metasurface for (a, d) $x$- and (b, e) $y$-polarized detection. (c, f) shows the intensity difference $\Delta \text{I} =\text{I}_y - \text{I}_x$, where $\text{I}_x$ and $\text{I}_y$ denote the $x$-polarized and $y$-polarized intensities, respectively. The dashed lines show the first-order solution of the grating equation at the respective wavelength for (a-c) $\lambda = \SI{590}{\nano\meter}$ and (d-f) $\lambda = \SI{610}{\nano\meter}$. The black circle in the center indicates a $\text{NA}=0.1$.}
	\label{fig:BFPImage}
\end{figure}

For further analysis of the polarization-dependent emission characteristics, we performed BFP imaging of fluorescence from the hybrid metasurface using the same setup and excitation light source as described for the momentum resolved spectroscopy measurements.
However, instead of projecting the BFP of the collection objective $(\text{NA}=0.9\,,100\times)$ onto the entrance slit of a spectrometer, it is directly projected onto an electron-multiplying charge-coupled device (EMCCD) camera.
Bandpass filters centered around \SI{610}{\nano\meter}  (\SI{590}{\nano\meter}) with a \SI{10}{\nano\meter} pass band are used to select the wavelength of the ED (MD) transition. 
Fig.~\ref{fig:BFPImage} shows corresponding BFP images for $x$- and $y$-polarized detection. 

Two intersecting bright features having the shape of sections of a circle appear as dominant features in the BFP images.
These correspond to the highly dispersive modes seen in the momentum-spectroscopy.
They appear both in $x$- and $y$-polarization, however, their dominant contribution is $x$-polarized.
By comparison with the grating equation (dashed lines), we can identify them as the lowest order diffractive modes of the underlying square lattice. The grating equations are detailed in S5.
Notably the solution of of the grating equation provides us with lattice periods of $p_x=\SI{220}{\nano\meter}$ and $p_y=\SI{440}{\nano\meter}$. This is different from our designed square unit cell with $p_x=p_y=\SI{440}{\nano\meter}$. However the unit cell consist of a dimer structure affectively reducing the $x$-period by a factor of two.
In order to discuss the spatially-resolved polarization dependency of the emission for the ED and MD transition induced by coupling to the dimers, we direct our attention to emission directions other than the diffractive modes.
No strongly localized features are observed for other directions, which is consistent with the low directionality associated with dimer nanoantennas.
Importantly, for the ED (MD), we find most of the numerical aperture outside of the diffractive features is dominated by $x$-polarized ($y$-polarized) light.
For example, for emission normally out of the sample plane, i.e., in the center of the BFP image ($\Gamma$-point), Fig.~\ref{fig:BFPImage} (c) shows that for the MD transition at $\lambda = \SI{590}{\nano\meter}$ the emission has a stronger $y$-polarized contribution with $\text{I}_y-\text{I}_x = 149$ counts at the $\Gamma$-point, while (f) reveals that for the ED transition at $\lambda = \SI{610}{\nano\meter}$ the emission has a stronger $x$-polarized contribution with $\text{I}_y-\text{I}_x = -385$ counts at the $\Gamma$-point. 

Altogether, we can conclude that the desired polarization dependence of emission from the ED and MD transition is observed over a broad range of emission angles.  In the difference maps (c, f) of S4, two unpolarized points appear at $[k_y/k_0 = 0, k_x/k_0 =\pm0.5]$, consistent with the expected signatures of a $C_2$-symmetric metasurface \cite{zhangChiralEmissionResonant2022}. With a slight tuning of the geometrical parameters  of the metasurface, mainly the period, these bands can be shifted to overlap at the $\Gamma$ point, which would result in a very strong $x$-polarized emission at normal incidence, as shown in the S4.

\section{Conclusion and Outlook}
In conclusion, we have experimentally investigated the polarization-dependent emission properties of a hybrid light-emitting metasurface consisting of a \ce{PMMA$:$EU(TTA)_3} coupled to a square array of all-dielectric elliptical a-Si:H dimers, which support polarized magnetic and electric modes.
In the gap between the two nanoresonators, a hot-spot of the electric (magnetic) field is observed for excitation with $x$-polarized ($y$-polarized) light at \SI{610}{\nano\meter} (\SI{590}{\nano\meter}).
By reciprocity, this system is expected to exhibit polarization-dependent emission behaviour, resulting from enhanced $x$-polarized emission for $x$-oriented EDs and enhanced $y$-polarized emission for $x$-oriented MDs.
We have experimentally confirmed this behavior. Channeling of light emitted via the electric dipole (ED) transition into $x$-polarized far-fields, and via the magnetic dipole (MD) transition into $y$-polarized far-fields, occurs over a wide range of emission angles, including both normal and off-normal directions.
Specifically, for $-0.3<k_{x}/k_{0}<0.3$ and $-0.1<k_{y}/k_{0}<0.1$, the ED emission is predominantly $x$-polarized while the MD emission is predominantly $y$-polarized.
The intensity ratios between the ED and MD transitions for $x$- and $y$-polarized emission is changed by approximately $+50\%$ ($-50\%$) as compared for the value of the \ce{PMMA$:$EU(TTA)_3} film on the bare substrate. In addition, we observe unpolarized points at $k_{x}/k_{0}=\pm0.5,k_{y}/k_{0}=0$, consistent with the behavior expected for $C_{2}$-symmetric metasurfaces.

Our results unlock polarization as a new degree of freedom in the selective interaction of light with complex emitters hosting both electric and magnetic dipolar transitions.
An optimization of the system performance may, e.g., be achieved by selective placement of the emitters in the dimer gap to reduce the background fluorescence from areas of the \ce{PMMA$:$EU(TTA)_3} film outside of the hotspot and may ultimately allow for effective polarization routing of the emission originating from electronic transitions of different multipolar order \cite{rikersPolarizationDependentEnhancement2025, rikersDeterministicFabricationFluorescent2025}.
Importantly, our system does not primarily rely on color routing, and instead modifies an MD transition with magnetic field enhancement.
Altogether, our work advances the understanding of the interaction of magnetic dipolar transitions with tailored nanostructures, with potential applications in light-emitting metasurfaces and quantum-nanophotonics.

\section{Methods}
\textbf{Optimization of metasurface:}
The metasurface geometry was optimized by performing a grid-based parameter sweep over $D_x$, $D_z$, and the aspect ratio $\alpha = D_y / D_x$, with the period $p = \SI{400}{\nano\meter}$ and feedgap $g = \SI{40}{\nano\meter}$ fixed due to fabrication constraints. No gradient-based or evolutionary optimization algorithm was used; instead, discrete parameter combinations were evaluated to identify the design maximizing the target figure of merit (FOM) while simultaneously ensuring high absolute field enhancements (particularly magnetic enhancement at $\lambda_1$, as a high FOM with low absolute enhancement would be impractical for applications).

The polarization contrast for a field $\mathbf{F} \in \{\mathbf{E}, \mathbf{H}\}$ at wavelength $\lambda$ is defined as the volume-averaged difference in normalized intensity between $y$- and $x$-polarized excitation:

\begin{equation}
    C^{\mathbf{F}}(\lambda) = \frac{1}{V} \iiint \left( \frac{|\mathbf{F}^{y-\text{pol}}(\lambda)|^2}{|\mathbf{F_0}^{y-\text{pol}}(\lambda)|^2} - \frac{|\mathbf{F}^{x-\text{pol}}(\lambda)|^2}{|\mathbf{F_0}^{x-\text{pol}}(\lambda)|^2} \right) d^3 r,
    \label{eq:polcontrast}
\end{equation}

where $\mathbf{F_0}$ is the field in a bare \ce{PMMA} reference of the same thickness and $V$ is the \ce{PMMA} volume. Positive $C^{\mathbf{F}}(\lambda)$ indicates stronger enhancement under $y$-polarized light and negative values indicate selectivity for $x$-polarized enhancement.

The overall figure of merit was constructed to capture dual-wavelength, dual-polarization selectivity:

\begin{equation}
    \text{FOM} = \max \left[ C^{\mathbf{H}}(\lambda_1) - C^{\mathbf{E}}(\lambda_2) \right].
    \label{eq:FOM}
\end{equation}

\begin{figure}[hbt!]
    \centering
    \includegraphics[width=0.9\linewidth]{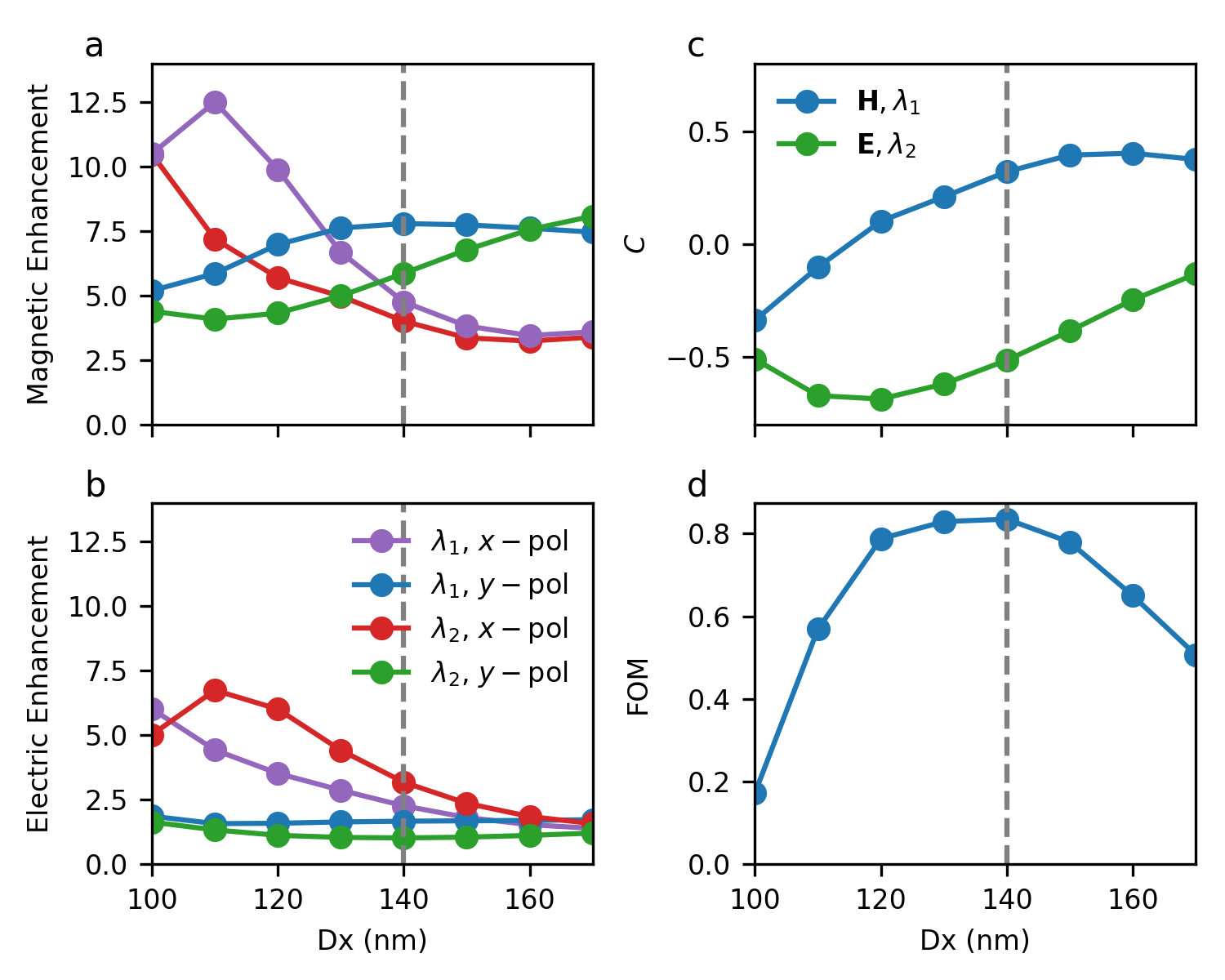}
    \caption{Parameter sweep over the dimer short axis $D_x$. (a,b) Field enhancement in the PMMA layer of the hybrid metasurface. (c) Polarization contrast for the magnetic field $\mathbf{H}$ at $\lambda_1 = \SI{590}{\nano\meter}$ and the electric field $\mathbf{E}$ at $\lambda_2 = \SI{610}{\nano\meter}$ (Equation~\ref{eq:polcontrast}). (d) Figure of merit (FOM) (Equation~\ref{eq:FOM}).}
    \label{fig:SimOpt}
\end{figure}

This expression rewards large positive magnetic contrast at $\lambda_1$ (strong $y$-polarized magnetic hotspot) while penalizing large positive electric contrast at $\lambda_2$ (unwanted $y$-polarized electric enhancement at the wavelength intended for $x$-polarized electric selectivity). The sweep results of magnetic and electric enhancements, individual polarization contrasts, and the resulting FOM, as a function of $D_x$ are presented in Fig.~\ref{fig:SimOpt}.

The parameter combination $\alpha = 1.5$ and $D_z = \SI{70}{\nano\meter}$ was identified as optimal from the initial sweep. A subsequent refinement over $D_x$ showed that $D_x = \SI{140}{\nano\meter}$ (hence $D_y = \SI{210}{\nano\meter}$) yields the highest FOM while also providing strong absolute magnetic field enhancement at $\lambda_1$ under $y$-polarized excitation and good electric enhancement at $\lambda_2$ under $x$-polarized excitation. 

\begin{acknowledgement}
This work was funded by the Deutsche Forschungsgemeinschaft (DFG, German Research Foundation) through the International Research Training Group (IRTG) 2675 “META-ACTIVE”, project number 437527638. This work used the ACT node of the NCRIS-enabled Australian National Fabrication Facility (ANFF-ACT). A. B. gratefully acknowledges financial support from Spanish national project No. PID2022-137857NA-I00. A. B. thanks MICINN for the Ramon y Cajal Fellowship (grant No. RYC2021-030880-I).
\end{acknowledgement}


\begin{suppinfo}
\begin{itemize}
    \item Description of sample fabrication process.
    \item Description of optical characterization setup.
    \item Refractive index of a-Si:H
    \item BFP images of sample with $p=\SI{400}{\nano\meter}$
    \item Grating equation
\end{itemize}
\end{suppinfo}

\bibliography{references.bib}

\end{document}